# Facile synthesis of CoSi alloy with rich vacancy for base- and solvent-free aerobic oxidation of aromatic alcohols

Zhiyue Zhao[a,†], Zhiwei Jiang[a,†], Yizhe Huang[a], Mebrouka Boubeche[b], Valentina G. Matveeva[c], Hector F. Garces[d], Huixia Luo[b], Kai Yan[a,e,*]

[a] China School of Environmental Science and Engineering, Sun Yat-Sen University, Guangzhou 510006, Guangdong, China

[b] School of Materials Science and Engineering, State Key Laboratory of Optoelectronic Materials and Technologies, Sun Yat-Sen University, Guangzhou 510006, Guangdong, China

[c] Department of Biotechnology and Chemistry, Tver State Technical University, Tver 170026, Tver, Russian

[d] School of Engineering, Brown University, Providence 02912, Rhode Island, United States

[e] Guangdong Laboratory for Lingnan Modern Agriculture, South China Agricultural University, Guangzhou 510642, Guangdong, China

**Abstract:** Rational design and green synthesis of low-cost and robust catalysts efficient for the selective oxidation of various alcohols are full of challenges. Herein, we report a fast and solvent-free arc-melting (AM) method to controllably synthesize semimetal CoSi alloy (abbreviated as AM-CoSi) that is efficient for the base- and solvent-free oxidation of six types of aromatic alcohols. X-ray absorption fine structure (XAFS), electron paramagnetic resonance (EPR), and aberration corrected high angle annular dark field scanning transmission electron microscope (AC HAADF-STEM) confirmed the successful synthesis of AM-CoSi with rich Si vacancy ($Si_v$). The as-prepared CoSi alloy catalysts exhibit an order of magnitude activity enhancement in the oxidation of model reactant benzyl alcohol (BAL) to benzyl benzoate (BBE) compared with its mono counterparts, whereas 70 % yield of BBE which is the highest yield to date. Experimental results and DFT calculations well verify that the CoSi alloy structure improves the BAL conversion and Si vacancy mainly contributes to the generation of BBE. After that, CoSi alloy maintains high stability and a potential pathway is rationally proposed. Besides, CoSi alloy also efficiently works for the selective oxidation of various alcohols with different groups. This work demonstrates for the first time that semimetal CoSi alloy is robust for the green oxidation of various alcohols and provides a vast opportunity for reasonable design and application of other semimetal alloy catalysts.

Received XXXX. Accepted XXXX. Published XXXX.

＊Corresponding author E-mail: yank9@mail.sysu.edu.cn

† Contributed equally to this work.

This work was supported by National Natural Science Foundation of China (21776324, 22078374), and the

# 1. Introduction

Selective oxidation of alcohols to synthesize aldehydes, acids and esters is extremely important in the synthesis of pharmaceuticals, perfumes, and fine chemicals [1-5]. Aromatic alcohols are common platform molecules that can be acquired from renewable biomass sources [6-8]. Tremendous efforts have been devoted to the oxidation of aromatic alcohols because the produced compounds can work as a versatile intermediate in organic synthesis and the manufacture of dyes, pharmaceuticals, perfumes, and preservatives [9-14]. Benzyl alcohol (BAL), the simplest aromatic alcohol, has been extensively studied as a representative model reactant. To date, many catalysts have been developed for the oxidation of BAL in the presence of alkaline environment or/and organic solvent [15, 16]. It has been proved that alkaline conditions can greatly improve catalytic activity due to their high hydrogen abstraction rate [17]. However, the alkaline agents easily lead to secondary environmental pollution issues, limiting their further applications. Consequently, a base-free oxidation system is more desired. Therefore, the establishment of a green catalytic system for base- and solvent-free oxidation of aromatic alcohols with superior performance are of great scientific interest.

Recently, heterogeneous catalysts based on noble-metal, such as Au- and Pd-derived catalysts including their alloys, have been developed for the solvent-free aerobic oxidation of BAL [18-21]. Among them, $Pd_3Pb$ alloy/$P_{25}$ catalyst was fabricated in the presence of organic compounds (oleic acid and octadecene) for the oxidation of BAL at 130 °C and 1 MPa $O_2$ for 1 h, wherein 91.3% conversion and 91% selectivity of benzaldehyde (BZH) was obtained [22]. Due to the high-cost and complex synthesis processes, noble-free metal oxide candidates have also been synthesized [23], especially Co-based catalysts. For instance, Beller et al. reported that $Co_3O_4$/NGr@C catalyst showed extraordinary performance (94% of yield) for the oxidative esterification of BAL [24]. Despite good progress reached, relatively low reaction activity and stability halted their practical applications. Recently, high entropy oxides (HEO) have attracted growing attention due to abundant cation compositions, thermal stability, and diverse active sites [25, 26]. A holey lamellar HEO ($Co_{0.2}Ni_{0.2}Cu_{0.2}Mg_{0.2}Zn_{0.2}O$) has been employed for the solvent-free oxidation of BAL with high catalytic activity and 67% selectivity of BZH or 62% of benzoic acid through adjusting the reaction conditions [27]. However, the fabrication of holey lamellar HEO catalyst involves the use of organic solvents, a complicated and time-consuming process, greatly hindering its wide utilization. Therefore, the development of a green and fast method to fabricate noble metal-free catalysts for base-free, even solvent-free aerobic oxidation of aromatic alcohols with high performance is highly desired. Furthermore, due to the specific electronic structures, high catalytic activities, and excellent stability in the harsh environment, transition metal silicides alloys have been obtained growing attention in catalytic applications

[28].

Herein, semimetal CoSi alloy with Co vacancy and Si vacancy was prepared through the fast and solvent-free arc-melting (AM) method. The fabricated AM-CoSi alloy catalyst was for the first time utilized for the base- and solvent-free aerobic oxidation of six aromatic alcohols. Aided by AC HAADF-STEM, XAFS, and EPR measurements, abundant Si vacancies and a few Co vacancies are formed in the as-prepared AM-CoSi alloy. The as-prepared CoSi alloy catalyst exhibits much-enhanced performance in the selective oxidation of BAL to BZH or BBE compared with its mono counterparts, whereas 70 % yield of BBE is the highest yield reported to date. In addition, the as-prepared AM-CoSi alloy also works universally for the highly selective oxidation of other aromatic alcohols with electron-donating and withdrawing groups (e.g., 2-methyl-, 3-methyl-, 4-methyl-, 4-methoxy-, and 4-nitrobenzyl alcohol). Furthermore, AM-CoSi catalyst maintains a highly stable structure and has excellent stability even over 5 runs. After this, the DFT theory further elucidates the potential pathway and is well matched with the experimental results, confirming Si vacancies and the alloy structure of AM-CoSi are beneficial for the superior activity and stability. This work provides a new strategy for the efficient upgrading of various alcohols by the new type of semimetal alloy candidates.

## 2. Experimental section

### 2.1. Chemicals

Cobalt powder (99.5%) was obtained from Macklin factory. Silicon powder (99.9%), benzyl alcohol ($C_7H_8O$, ≥99%), 2-methylbenzaldehyde ($C_8H_8O$, 99%), 3-methylbenzaldehyde ($C_8H_8O$, 98%), 4-methylbenzyl alcohol ($C_8H_{10}O$, ≥99%), 4-methylbenzaldehyde ($C_8H_8O$, 98%), 4-methoxybenzyl alcohol ($C_8H_{10}O_2$, 98%), 4-methoxybenzaldehyde ($C_8H_8O_2$, 98%), benzoic acid ($C_7H_6O_2$, 98%), 2-methylbenzoic acid ($C_8H_8O_2$, 98%), 3-methylbenzoic acid ($C_7H_6O_2$, 99%), 4-methylbenzoic acid ($C_7H_6O_2$, 98%), 4-nitrobenzyl alcohol ($C_7H_7NO_3$, 98%) and 4-nitrobenzaldehyde ($C_7H_5NO_3$, 98%) were purchased from Aladdin factory. 2-Methylbenzyl alcohol ($C_8H_{10}O$, 98%) was purchased from Shanghai Acmec Biochemical Co., Ltd and 3-methylbenzyl alcohol ($C_8H_{10}O$, ≥99%) was acquired from Energy Chemical.

### 2.2. Preparation of CoSi catalysts

AM-CoSi alloy catalyst was rapidly synthesized by arc melting stoichiometric mixtures of silicon and Co metals on a metal melting furnace (MSM20-7) under argon atmosphere equipped with a water-cooled copper hearth. Typically, 96.8 mg of Si powders and 203.2 mg of Co powders were weighed at a molar ratio of 1:1, and then ground and thoroughly mixed. The ground powders were pressed by hydraulic press, and then transferred to a metal melting furnace to melt into silvery grey metal ingots within 5 minutes. The obtained ingots were ground into powders before using as the catalyst.

S-CoSi catalyst was synthesized through the seal tube calcination method. Typically, 203.2 mg Co and 96.8 mg Si were weighed in a molar ratio of 1:1, ground and thoroughly mixed. After grinding, the sample was sealed into a vacuum quartz tube at 1000 °C with a heating rate of 1 °C /min for 48 hours and then cooled naturally to achieve pure phase CoSi ingots, denoted S-CoSi. The obtained S-CoSi ingots were ground into powders before using as the catalyst.

**2.3. Catalyst characterization**

Transmission electron microscopy (TEM) analysis was carried out on Fei TECNAI G2 F20 field emission transmission electron microscope, which operates at 200 kV accelerating voltage. High resolution TEM (HR-TEM), high angle annular dark field (HAADF) scanning transmission electron microscope (STEM) and energy dispersive X-ray energy dispersive spectroscopy (EDS) were performed on the aberration corrected scanning transmission electron microscope (jemarm300f-grand-arm) at 200 kV. Aberration corrected high angle annular dark field scanning transmission electron microscope (AC HAADF-STEM) images were collected in Themis Z single spherical aberration corrected Transmission Electron Microscope (ACTEM) operating at 300 kV. Routine test steps of TEM: a certain amount of sample is weighed and dispersed in ethanol, dispersed by ultrasonic for 5 to 10 mins, and dropped on the copper wire for detection. The sample XRD patterns were acquired using the Rigaku Ultima IV X-ray diffractometer with Cu-$K_\alpha$ radiation at 40 mA with 40 kV. The scanning rate is 10 °/min, and the range of 2-theta was 10 – 90°. The XPS spectra of the samples were collected by Thermo Scientific model 250Xi spectrometer using a monochromated Al $K\alpha$ radiation. The binding energies were referenced to the C 1$s$ line at 284.8 eV. X-band electron paramagnetic resonance (EPR) was conducted by Brooke a300. Inductively coupled plasma-optical emission spectrometry (ICP-OES) experiments were measured on Agilent 5110. $O_2$-temperature programmed desorption (TPD), $CO_2$-TPD experiments were carried out on a Quantachrome chemisorption analyzer equipped with a thermal conductivity detector.

The X-ray absorption fine structure (XAFS) spectra Co K-edge and Si K-edge were collected at BL14W1 beamline of Shanghai Synchrotron Radiation Facility (SSRF). The data were collected in fluorescence mode using a Lytle detector while the corresponding reference samples were collected in transmission mode. The sample was ground and uniformly daubed on the special adhesive tape. The acquired XAFS data were processed according to the standard procedures using the ATHENA module of Demeter software packages. The Extended x-ray absorption fine structure (EXAFS) spectra were obtained by subtracting the post-edge background from the overall absorption and then normalizing with respect to the edge-jump step. Subsequently, the χ(k) data of Fourier transformed to real (R) space using a hanning windows (dk = 1.0 Å$^{-1}$) to separate the EXAFS contributions from different coordination shells.

## 2.4. Catalytic activity test

The base- and solvent-free aerobic oxidation of aromatic alcohols (benzyl alcohol, 2-methylbenzyl alcohol, 3-methylbenzyl alcohol, 4-methylbenzyl alcohol, 4-methoxybenzyl alcohol and 4-nitrobenzyl alcohol) were carried out in a 10 mL stainless steel reactor under magnetic stirring (PXGF-H8-25, Xi'an Taikang Biotechnology Co., Ltd). In a typical reaction, 3 mL of benzyl alcohol and 10 mg of catalyst are added into the reactor. The reactor was flashed with 1 MPa of $O_2$ three times to remove the air, and then reacted at the desired temperature for the given time and oxygen pressure. After the reaction, the product was diluted 100 times with ethanol and then ultrasonicated for 15 mins. After separating the solid catalyst, the liquid products were analyzed by gas chromatography (GC, Techcomp 790). Solvent-free selective oxidation of 2-methylbenzyl alcohol, 3-methylbenzyl alcohol and 4-methylbenzyl alcohol were also conducted under the similar conditions.

BAL conversion, BZH, BAD and BBE selectivity were calculated according to Equations (1) to (2):

$$\text{Conversion (\%)} = (C_0 - C_A)/C_0 \times 100\% \qquad (1)$$

$$\text{Selectivity (\%)} = C_K/(C_0 - C_A) \times 100\% \qquad (2)$$

where $C_0$ was the initial concentration of the reactant, and $C_A$ was the concentration after the reactant reaction, and $C_K$ was the concentration of the product (benzaldehyde, benzoic acid, or benzyl benzoate).

## 2.5. Theoretical calculation

We have employed the Vienna Ab Initio Package (VASP) [29, 30] to perform all the density functional theory (DFT) calculations within the generalized gradient approximation (GGA) using the PBE [31] formulation. We have chosen the projected augmented wave (PAW) potentials [32, 33] to describe the ionic cores and take valence electrons into account using a plane wave basis set with a kinetic energy cutoff of 400 eV. Partial occupancies of the Kohn−Sham orbitals were allowed using the Gaussian smearing method and a width of 0.05 eV. The electronic energy was considered self-consistent when the energy change was smaller than $10^{-5}$ eV. A geometry optimization was considered convergent when the force change was smaller than 0.02 eV/Å. Grimme's DFT-D3 methodology [34] was used to describe the dispersion interactions.

The equilibrium lattice constant of cubic tetraauricupride-structured CoSi unit cell was optimized, when using a 11 × 11 × 11 Monkhorst-Pack k-point grid for Brillouin zone sampling, to be a = 4.390 Å. We then use it to construct a CoSi (200) surface model with p (3 × 3) periodicity in the x and y directions and 2 stoichiometric layers in the z direction separated by a vacuum layer in the depth of 15 Å to separate the surface slab from its periodic duplicates. This surface model contains 48 Co and 48 Si atoms. During structural optimizations, the gamma point in the Brillouin zone was used for *k*-point sampling, and the bottom stoichiometric layer was fixed while the top one was allowed to relax.

The reaction energy ($E_{reaction}$) in the reaction pathway was calculated by Equation (3):

$$E_{reaction} = \sum_{i=1}^{n} E_{adsorbate(i)} - nE_{surface} - 2E_{PHCH_2OH_{(g)}} - E_{O_2(g)} \qquad (3)$$

where n is the number of adsorbates, $E_{adsorbate}$ is the total energy of the surface with an adsorbate, $E_{surface}$ is the total energy of the surface without adsorbate, $E_{PHCH2OH(g)}$ is the energy of the gas phase PhCH$_2$OH and $E_{O2(g)}$ is the energy of the gas phase O$_2$.

## 3. Results and discussion

### 3.1. Synthesis and characterization

CoSi alloy is facilely and fast synthesized through the solvent-free arc-melting method in 5 mins as depicted in **Fig. 1**a. The crystal structure of the fabricated AM-CoSi sample was firstly confirmed by powder X-ray diffraction (XRD). Peaks (Fig. S1) located at 28.4°, 34.9°, 40.5°, 45.6°, 50.2°, 62.6°, 70.2°, and 77.3° correspond to the (110), (111), (200), (210), (211), (221), (311) and (320) facets of CoSi alloy (JCPDS No. 50-1337) [35]. No impure phase was observed, confirming the high purity and single phase of the alloyed structure. The Co/Si ratio determined by inductively coupled plasma atomic emission spectroscopy (ICP-AES) was 1.08, which was close to the initial stoichiometric value and indicated part loss during the synthesis.

The textural information of AM-CoSi alloy was then examined by transmission electron microscopy (TEM). TEM images show the irregular microscale morphology in low resolution (Fig. S2a). In addition, four diffraction planes (110), (111), (210), and (220) are well assigned in the selected area electron diffraction (SAED) pattern of AM-CoSi sample (Inset in Fig. S2a), further indicating the successful synthesis of alloy. Fig. 1b and Figs. S2b–S2d further show high-resolution transmission electron microscopy (HR-TEM) images of AM-CoSi. The observed *d*-space of 0.225 nm is attributed to the (200) lattice plane (Inset of Fig. 1b) and the clear layer laminated structure of AM-CoSi is clearly displayed. Besides, the clear dots (Fig. 1c and Fig. S2d) confirm the existence of surface defects. Line blemish (Fig. 1d) shows the edge dislocation of Co or Si, resulting in line defects. Plane defects (Fig. 1e) exhibit the small angle of 8° grain boundaries, indicating the reconfiguration of AM-CoSi alloy structure. Furthermore, the d-spacing of 0.220 nm assigned to the CoSi alloy (200) plane in Fig. 1d is smaller than that of 0.225 nm in Fig. 1b, confirming the presence of compressive strain due to the shift of d-band [36]. The defect is thus likely provoked by the compressive strain of Co-Si lattice [37]. Besides, Co and Si elements are finely distributed in the AM-CoSi alloy as confirmed by high-angle annular dark-field scanning transmission electron microscopy (HAADF-STEM). Spherical aberration-corrected TEM (ACTEM) is employed to analyze the atom distributions in AM-CoSi alloy, as shown in Fig. 1f, and Figs. S2g-S2i, respectively. The AC HAADF-STEM image (Fig. 1f) displays the general order arrangement of Co and Si atoms with space group

*P*213. According to the crystal parameter of cubic CoSi, the distance between Co-Co was larger than that of two adjacent Si atoms. Some defects also clearly appear on the (200) plane (Fig. 1g). The production of vacancies in AM-CoSi might be due to the rapid cooling during preparation process [38]. To further identify the type of vacancies in the synthesized AM-CoSi alloy, electron paramagnetic resonance (EPR) was carried out. As shown in Fig. S3a, a strong EPR signal ($g = 2.002$) and a weak EPR signal ($g = 2.007$) are attributed to Si vacancies ($Si_v$) and Co vacancies ($Co_v$), respectively, suggesting the divacancy existed and a little higher $Si_v$ concentration in AM-CoSi alloy [39]. Based on these results, the schematic configuration of AM-CoSi alloy is then schematically illustrated in Fig. S3b.

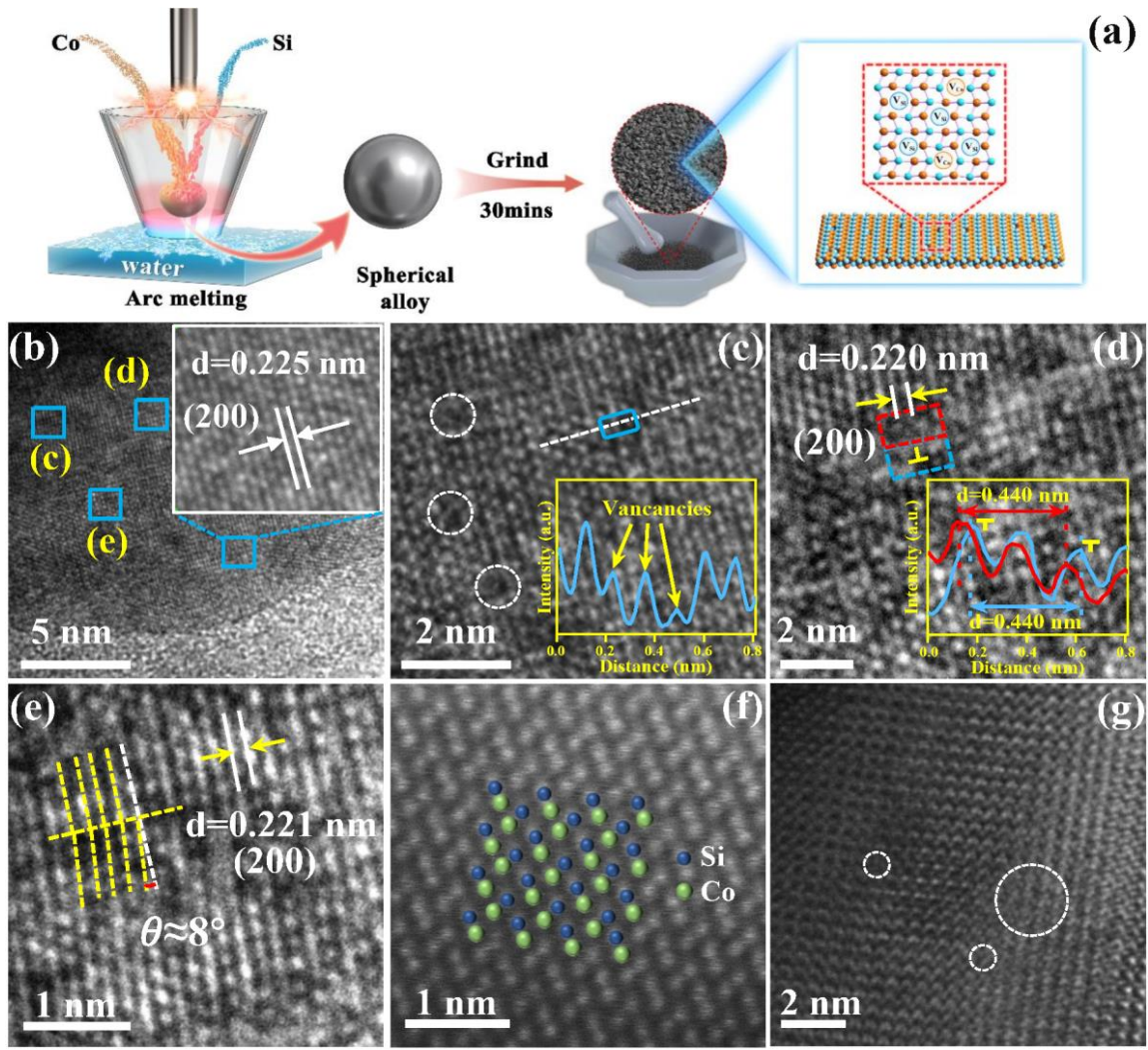

**Fig. 1.** (a) Schematic representation of the synthesis of AM-CoSi alloy. (b) HR-TEM images of AM-CoSi alloy. (c) Point defects (inset shows the corresponding image intensity line profiles). (d) Line defects (inset shows the corresponding image intensity line profiles). (e) Plane defects. (f) AC HAADF-STEM image and (g) ACTEM image of AM-CoSi alloy.

AM-CoSi alloy is then characterized by X-ray adsorption fine structure spectroscopy (XAFS) to investigate the local structure and coordination environment of the Co atom (**Fig. 2**) [40]. Fig. 2a displays the Co K-edge of the X-ray absorption near-edge structure (XANES) spectra of AM-CoSi, $Co_3O_4$, Co foil, and standard CoSi sample (abbreviated as S-CoSi). The Co K-edge of AM-CoSi and S-CoSi is very similar to that of Co foil (inset in Fig. 2a), indicating that the valence state of Co in AM-CoSi was close to zero. Fourier-transformed $k^3$-weighted EXAFS spectra at the Co edge are shown in Fig. 2b. The S-CoSi exhibited two major obvious peaks attributed to the shell of Co-Si (1.89 Å) and Co-Co (2.71 Å), respectively. Compared with S-CoSi, the intensity of Co-Si as well as Co-Co shell of AM-CoSi alloy decreased and the Co-Si peak shifted to a longer distance, which was possibly due to the d-orbit shift [37]. As shown in Fig. S4, the Co K-edge EXAFS $k^3\chi(k)$ oscillation curve for AM-CoSi alloy is similar to that of the S-CoSi, indicating that the ab-plane structure is well maintained. The wavelet transform (WT) contour plots of AM-CoSi, $Co_3O_4$, Co foil, and S-CoSi are display in Fig. 2c and Fig. S5, respectively. For AM-CoSi, $k$ values were located at ~ 5.59 and 11.07 Å$^{-1}$, corresponding to the Co-Si and Co-Co bond, respectively. Fitted EXAFS data of AM-CoSi alloy reveal that the Co-Si coordination number and the Co-Co coordination number are 6.5 and 2.5 on average (Fig. 2d, Table S1), respectively, which are both smaller than those of the S-CoSi alloy (Fig. S6), further revealing the co-existence of Si and Co vacancies. This work will focus on the effects of vacancies and the CoSi structure for BAL selective oxidation and AM-CoSi and S-CoSi were used as catalysts.

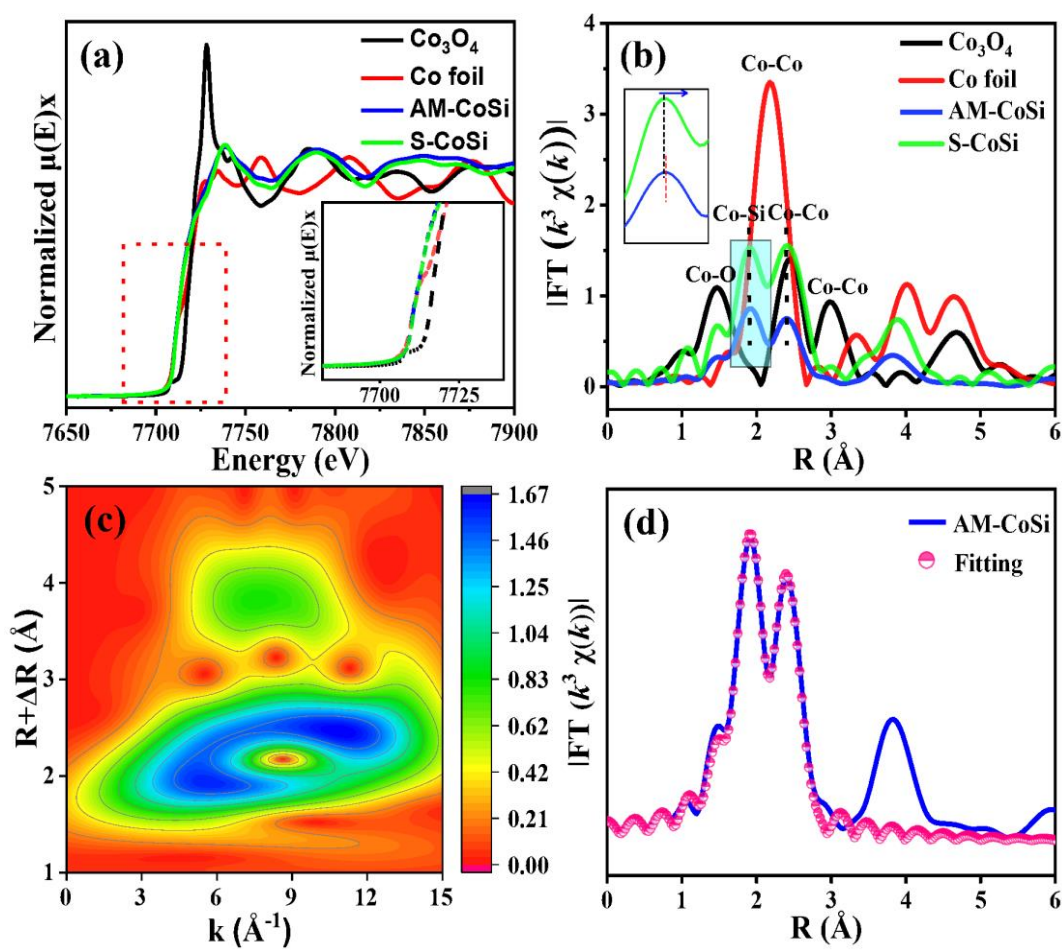

**Fig. 2.** Structural characterization by XAFS spectroscopy: (a) Experimental Co K-edge XANES spectra and (b) Fourier-transformed magnitude of the experimental Co K-edge EXAFS signal of AM-CoSi alloy and reference samples (Co foil, $Co_3O_4$, and S-CoSi). (c) Wavelet transform (WT) for the $k^3$-weighted EXAFS signals of AM-CoSi alloy. (d) Corresponding EXAFS R-space fitting curve of AM-CoSi alloy.

## 3.2. Catalytic oxidative activity

To evaluate the catalytic performance of AM-CoSi alloy, several aromatic alcohols are investigated in the base- and solvent-free oxidation system. BAL is firstly chosen as the reaction model (**Fig. 3**a). Three products, BZH, BAD, and BBE, are mainly produced in the oxidation process under 5 bar $O_2$ for 6 h. With the increase of reaction temperature from 100 °C to 220 °C, the conversion of BAL sharply increases from 3.5% to 54.8% at 140 °C and then slowly up to 64.5% at 220 °C (Table S2, entries 1-7). The selectivity of BBE also rapidly improves from 4.7% at 140 °C to 37.5% at 220 °C, revealing that high temperature promotes the generation of BBE. Reaction time on the selectivity of BZH and BBE at different temperature are then further evaluated. Fig. 3b shows the effect of reaction time on the yield of BZH at 140 °C under 5 bar $O_2$. With an increase in reaction time, the conversion of BAL and the selectivity of BAD as well as BBE are sharply enhanced, while the selectivity of BZH is reduced. A similar trend for the yield of BBE over AM-CoSi alloy is observed at 220 °C (Fig. 3c). Influence of reaction pressure on the yield of BZH (at 140 °C for 6 h) and BBE (at 220 °C for 24 h) is also investigated (Table S2, entries 12-16, and 20-23), as shown in Figs. 3d and 3e. It is clear to see the increased reaction pressure would promote the conversion of BAL and the selectivity of BAD and BBE. When the $O_2$ pressure increases from 3 to 9 bar at 220 °C, the selectivity of BZH is reduced by 10.7%, while the selectivity of BBE has an increase of 37.8% to 70%. By adjusting the catalytic reaction conditions, 70% selectivity to BBE with up to 99.9% conversion of BAL can be achieved at 220 °C under 9 bar $O_2$ for 24 h. On the other hand, 84.7% selectivity to BZH at 54.8% conversion of BAL can be obtained at 140 °C under 5 bar $O_2$ for 6 h. To the best of our knowledge, 70% yield of BBE is the highest value reported so far. The gas phase product was also detected and $CO_2$ was observed as a coproduct. Besides, the activity for the synthesis BZH is also comparable with those previously reported catalysts that include noble-metal catalysts as depicted in Table S3. The above results revealed that low temperature (< 140 °C) is beneficial to the formation of BZH but high temperature improve the yield of BBE during BAL oxidation under the base- and solvent-free conditions. Furthermore, the BAL conversion is relatively low at low temperature.

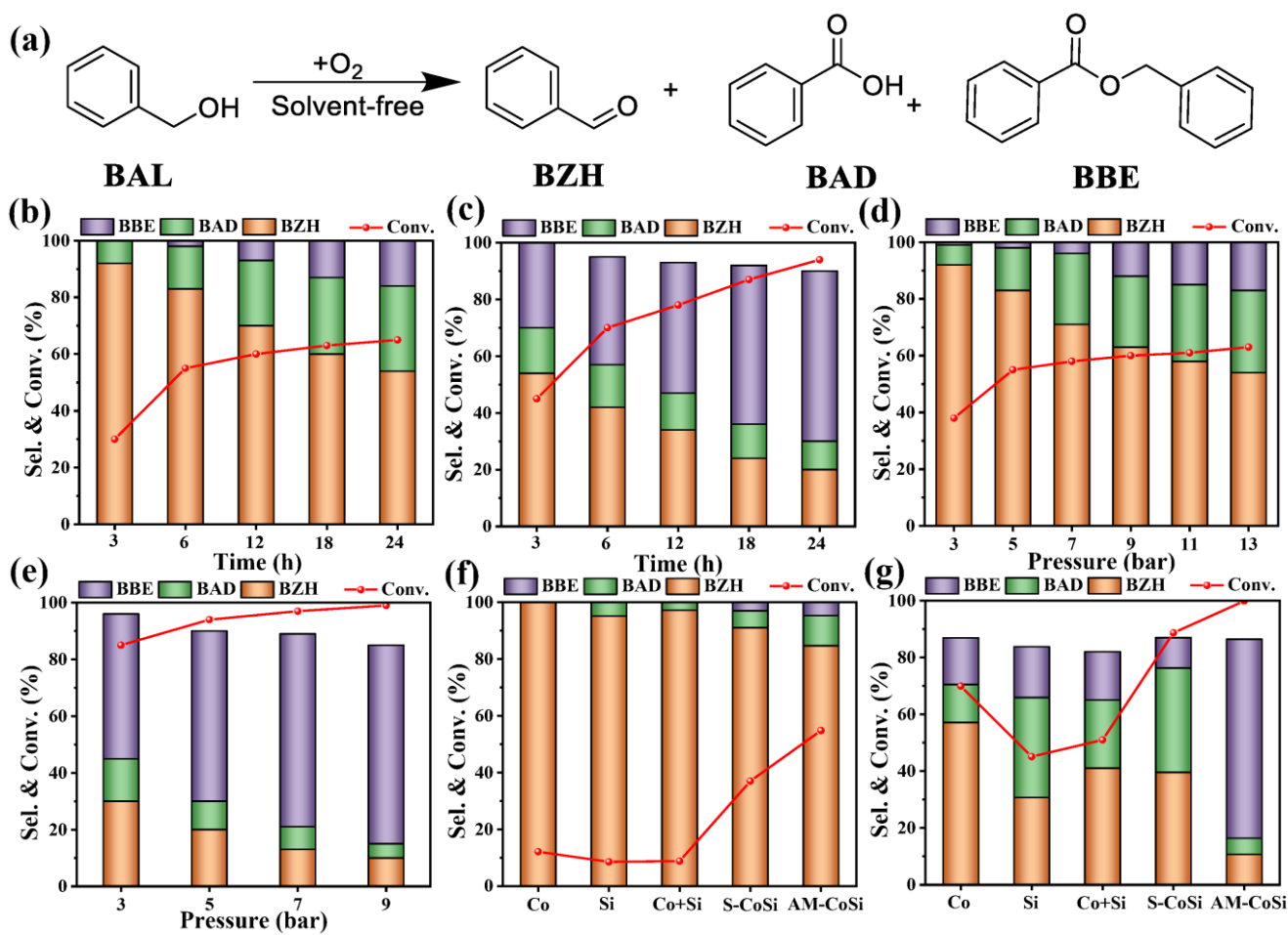

**Fig. 3.** (a) Selective oxidation diagram of BAL over AM-CoSi catalyst. The effect of time for the BAL solvent-free aerobic oxidation over AM-CoSi at 140 °C (b) and at 220 °C (c) under 5 bar $O_2$, respectively. The effect of $O_2$ pressure for the BAL solvent-free aerobic oxidation over AM-CoSi at 140 °C for 6 h (d) and 220 °C for 24 h (e). Product distributions of the oxidation of BAL over different catalysts at 140 °C for 6 h under 5 bar $O_2$ (f) and at 220 °C for 24 h under 9 bar $O_2$ (g). Reaction conditions: 3 mL of BAL, 10 mg of catalyst.

To better understand why AM-CoSi catalyst is good for the oxidation of BAL, the oxygen transportation capacity of AM-CoSi alloy catalysts is then investigated by $O_2$-temperature programmed desorption ($O_2$-TPD) as shown in Fig. S7. The desorption peaks located at ~200 °C, 300 – 500 °C, and >500 °C are assigned to molecular oxygen species adsorbed on vacancy, atomic state oxygen and crystal lattice oxygen, respectively [41, 42]. According to the $O_2$-TPD curve of AM-CoSi, the strong broad peak is attributed to atomic state oxygen desorption, demonstrating the high oxygen absorbing capacity of CoSi alloy. In addition, $CO_2$-TPD test is then employed to investigate the potential basicity of AM-CoSi catalyst, as shown in Fig. S8. A strong peak at 333 °C is observed in $CO_2$-TPD curve, manifesting a moderate basicity existed.[43] The moderate basicity is beneficial for the selective oxidation of alcohols.[44]

To further understand the role of AM-CoSi catalyst, AM-CoSi alloys are compared with their monometallic Co, counterpart Si, and standard CoSi alloy (S-CoSi) for the base- and solvent-free oxidation of aromatic alcohols (Table S4). Fig. 3f shows the product distributions over different catalysts at 140 °C for 6 h. The individually mono-Co and Si powder catalysts both show low catalytic activities with conversions of only 12.1% and 8.5%, respectively (Table S4, entries 1-2). The physically mixed catalyst (a mixture of Co and Si powders) displays no improvement with only 8.8% conversion (Table S4, entry 3). S-CoSi alloy is also used to oxidize BAL at 140 °C for 6 h and exhibited a high conversion of 37% (Table S4, entry 4). These results demonstrated that the alloyed structure was important for the BAL conversion. However, the activity of S-CoSi is much lower than that of AM-CoSi, which is likely due to the rich vacancy in AM-CoSi. To study the role of Co or Si to produce BBE, Co and Si powder and their mixture catalysts are also used to oxidize BAL at 220 °C for 24 h under 9 bar $O_2$ (Fig. 3g). Relatively, 69.9% conversion with 11.5% yield of BBE is achieved over Co powder (Table S4, entry 5) while 45.1% conversion with 8.1% yield of BBE is obtained over Si powder (Table S4, entry 6), demonstrating that Co is beneficial for the oxidation of BAL at high temperature. The mixed catalyst exhibits a conversion of 51.3% with 8.8% yield of BBE (Table S4, entry 7). S-CoSi is also studied to oxidize BAL at 220 °C for 24 h under 9 bar $O_2$ (Table S4, entry 8). High conversion of BAL (88.7%) with 9.5% yield of BBE is achieved, which is much lower than the yield of BBE (70%) over AM-CoSi alloy under the same reaction condition, further suggesting that the alloy structure and vacancy in AM-CoSi are vital to the superior performance in the oxidation of BAL. Furthermore, the turnover frequency (TOF) of BAL oxidation over CoSi alloy and monometallic catalysts were also inverstigated, as shown in Table S4, and the TOF of CoSi alloy (10.5 $h^{-1}$ for AM-CoSi and 9.3 $h^{-1}$ for S-CoSi) is much larger than that of Si (1.5 $h^{-1}$), Co (5 $h^{-1}$), and Si/Co mixture (2.7 $h^{-1}$).

**Table 1.** Solvent-free aerobic oxidation of various alcohols over AM-CoSi catalyst.[a]

| Entry | Substrate | Main product | Conv. (%) | Sel. (%) | Yield (%) |
|---|---|---|---|---|---|
| 1 | 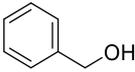 | 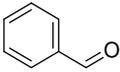 | 54.8 | 84.7 | 46.4 |
| 2 | 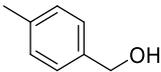 | 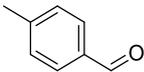 | 52.6 | 97.3 | 51.2 |
| 3 | 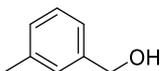 | 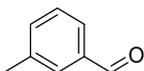 | 42.5 | 88.3 | 37.5 |
| 4 | 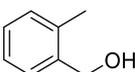 | 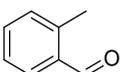 | 30.2 | 92.5 | 27.9 |
| 5 | 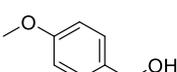 | 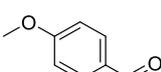 | 54.3 | 95.4 | 51.8 |
| 6 | 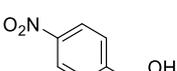 | 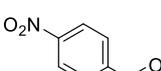 | 32.2 | 99.6 | 32.1 |

[a] Reaction conditions: 3 mL of the substrate, 10 mg of catalyst, 5 bar $O_2$, 140 °C, 6 h.

To evaluate the universal application of AM-CoSi alloy catalyst, a series of aromatic alcohols such as 2-, 3-, 4-methylbenzyl alcohol, 4-methoxybenzyl alcohol and 4-nitrobenzyl alcohol are chosen in the base- and solvent-free system. It is worth noting that the selectivity towards the aldehyde products is around 90% at 140 °C for 6 h, indicating that AM-CoSi alloy catalyst exhibits a high catalytic performance for the oxidation of various aromatic alcohols. For methylbenzyl alcohol with different substituent positions (Table 1), yields of the corresponding aldehydes are gradually decreased in the order of 4-methylbenzyl alcohol (52.6% conversion and 97.3% selectivity, Table 1, entry 2) > 3-methylbenzyl alcohol (42.5% conversion and 88.3% selectivity, Table 1, entry 3) > 2-methylbenzyl alcohol (30.2% conversion and 92.5% selectivity, Table 1, entry 4), showing that the position of the substituents clearly affects the oxidation reaction. The effect of substituent groups (electron donating and withdrawing groups) is also examined and compared. For aromatic with electron-donating groups (4-methoxybenzyl alcohol), the conversion essentially stays unchanged (~ 50%) with high selectivity to the corresponding aldehydes (Table 1, entry 5). For 4-nitrobenzyl alcohol with a withdrawing group, the catalytic activity has a certain reduction with 32.2% conversion (Table 1, entry 6), indicating that the electronic properties of the substituent groups also affected the catalytic activity. However, the catalytic oxidation activity for 4-nitrobenzyl alcohol over AM-CoSi alloy is much larger than that of the holey lamellar HEO or noble-metal catalysts [27].

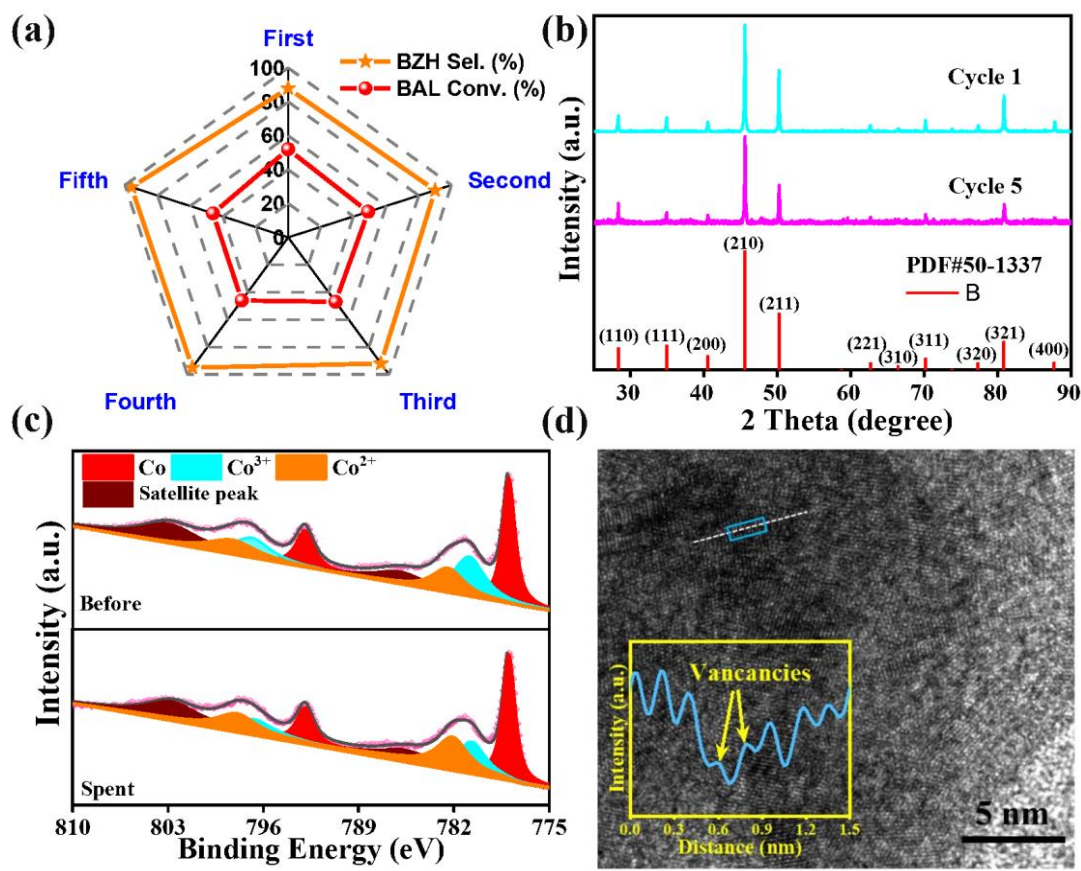

**Fig. 4.** (a) Reusability of AM-CoSi alloy catalyst for 5 runs. XRD patterns (b) and Co 2$p$ XPS spectra (c) of AM-CoSi before and after the reaction. (d) HR-TEM image of the spent AM-CoSi alloy catalyst (inset shows the corresponding image intensity line profiles along the white dotted line).

Catalyst stability is a crucial parameter for practical use. The spent AM-CoSi alloy catalyst is recovered and the catalytic activities of the spent AM-CoSi catalysts are shown in **Fig. 4**a. Even after 5 cycles, the selectivity and conversion are unchanged, showing high reaction stability of the catalyst. XRD patterns and high-resolution XPS spectra of the fresh and spent AM-CoSi alloy catalyst after 5 cycles are then carried out as shown in Figs. 4b and 4c, respectively. No clear change in XRD patterns of AM-CoSi before and after the reaction is observed (Fig. 4b). These results further confirmed that AM-CoSi alloy catalysts were stable under the reaction conditions. The XPS full spectra exhibit similar peaks between the fresh and spent catalyst in Fig. S9 and the data are listed in Table S5. Table S5 also summarizes the binding energies of AM-CoSi catalysts. The binding energies at 778.1 eV and 792.8 eV (Fig. 4c) correspond to the 2$p_{3/2}$ and 2$p_{1/2}$ of $Co^0$ in the spent AM-CoSi catalyst, respectively, which are similar to the fresh AM-CoSi alloy catalyst, confirming the good stability of AM-CoSi. Moreover, the oxidized Co species are also observed in the Co XPS spectra of fresh and spent AM-CoSi catalyst after the reaction, revealing that Co species in CoSi could be likely oxidized slightly during the oxidation reaction. However, only zero-valent Co species were obtained based on the XAFS analysis of AM-CoSi, which could be due to the fact that XAFS gives bulk information and XPS reflects the surface information. These information further confirm that the reaction is ocuurred on the surface of AM-CoSi catalyst. O 1$s$ (Fig. S10) of the spent AM-CoSi alloy catalyst located at 532.2 eV, which is also similar to the fresh AM-CoSi alloy. Besides, the peak at 532.2 eV is attributed to surface-absorbed oxygen in the oxidation process. The HR-TEM image of the spent AM-CoSi alloy is displayed in Fig. 4d. The image intensity alone white dotted line of Fig. 4d clearly shows lattice fringes, confirming that surface defects are also existed in the spent AM-CoSi alloy. In general, these results confirmed the CoSi alloy maintained high stability in the oxidation process.

### 3.3. Insights of reaction mechanism

To get more insights into the oxidation pathways of BAL, several control experiments are performed at 220 °C. Using BZH over AM-CoSi alloy catalyst in the presence of $O_2$ (5 bar) only lead to the benzoic acid (BAD) formation with 35.7% yield (Table S6, entry 1). However, no product is obtained without $O_2$ (Table S6, entry 2), which rules out the Tish-chenko mechanism in the base- and solvent-free oxidation of BAL (Scheme S1). When the mixture of BAL and BZH (the volume ratio of 1:1) is used as the reactant at 5 bar $O_2$, BBE is formed with 27.4% yield at 98.7% conversion of BAL (Table S6, entry 3). In comparison, no BBE is formed under $N_2$ (Table S6, entry 4), precluding the Cannizzaro pathway (Scheme S1). Further test using BAL and BAD (the mass ratio of 5:1) as a reactant in the presence of $N_2$ with and without AM-CoSi catalyst, BBE is selectively produced through a direct dehydrative esterification process (Table S6, entries 5-6). Furthermore, BAD is fully converted to BBE using AM-CoSi alloy while only 40.3% BAD is converted without a catalyst. These catalytic results further reveal that BBE is obtained through the direct esterification of BAL with BAD in the reaction process. Besides, we also examine the role of BZH in the esterification reaction between BAL and BAD. The addition of BZH would rapidly reduce the conversion of BAD (only 65.4%) and the formation of BBE with 62.5% yield (Table S6, entry 7), confirming that BZH restrains the esterification reaction and BBE is formed through the esterification of BAL with BAD. Based on the results of oxidation pathways of BAL and the conversion of BAL at different temperatures, high temperature promoted the obtained BZH to further oxide into BAD and followed by esterification over AM-CoSi, causing the continuous generation of BBE and a high yield of BBE. It was very difficult for BZH to further oxide at low temperature, leading to a high selectivity of BZH.

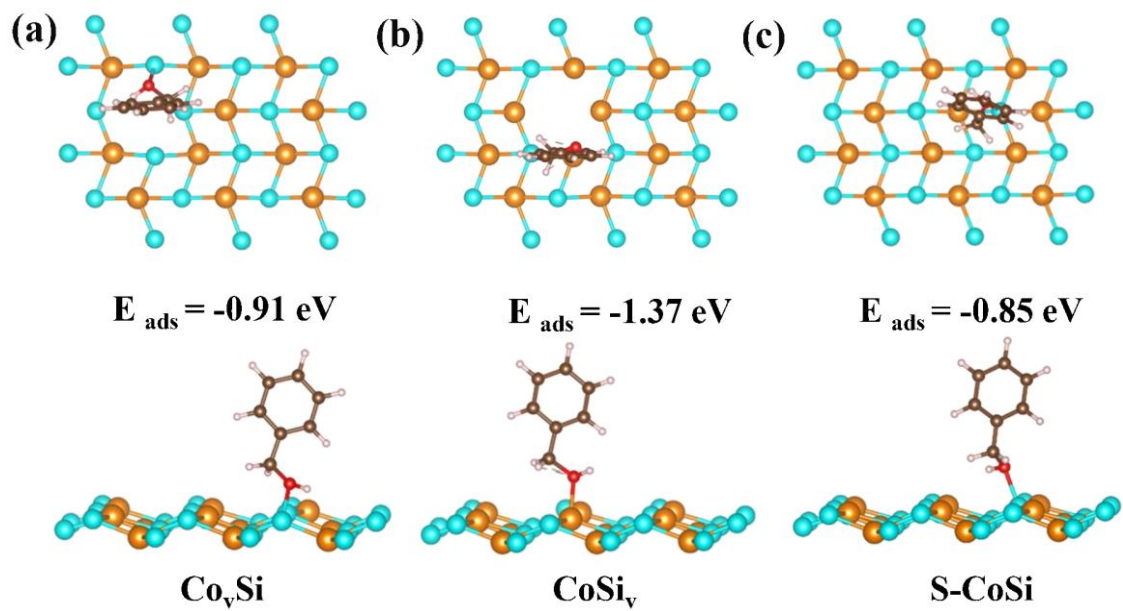

**Fig. 5.** DFT calculated conformation of adsorption states of BAL on (a) $Co_vSi$, (b) $CoSi_v$, and (c) S-CoSi.

To understand the reaction pathway and the role of CoSi alloy in the oxidation of BAL, DFT calculations are adopted to obtain the energies of reactants (BAL and BZH), intermediates, and products (BAD and BBE) on CoSi without vacancies (S-CoSi), CoSi with $Co_v$ ($Co_vSi$), and CoSi with $Si_v$ ($CoSi_v$) [29, 32, 45-47]. The configurations of BAL, BZH, BAD, and BBE with the lowest adsorption energies are shown in **Fig. 5** and Figs. S11-S13, and the adsorption energies are listed in Table S7, respectively. The Si deficiency leads to more Co sites exposed on the $CoSi_v$ surface, while the Co deficiency leads to more Si sites for the $Co_vSi$ surface. Therefore, due to the different exposed sites and coordination, the catalytic activities can be quite different on S-CoSi, $Co_vSi$, and $CoSi_v$ surfaces. The adsorption energies ($E_{ads}$) of the two considered reactants BAL and BZH on $CoSi_v$ are lower than those on S-CoSi and $Co_vSi$ (Table S7), demonstrating that Co sites near Si vacancies bind (stronger than various sites among the three surfaces) the oxygen atoms of hydroxyl groups in BAL or BZH. The reactants are better adsorbed on the $CoSi_v$ surfaces, such that the subsequent catalytic reaction is more likely to take place on the $CoSi_v$ surface. Besides, $E_{ads}$ of oxygen on $CoSi_v$ surfaces (-8.15eV) is lower than that on S-CoSi (-7.42eV) and $Co_vSi$ (-7.43eV) in table S8 and table S9, confirming that $CoSi_v$ showed a better oxygen activation.

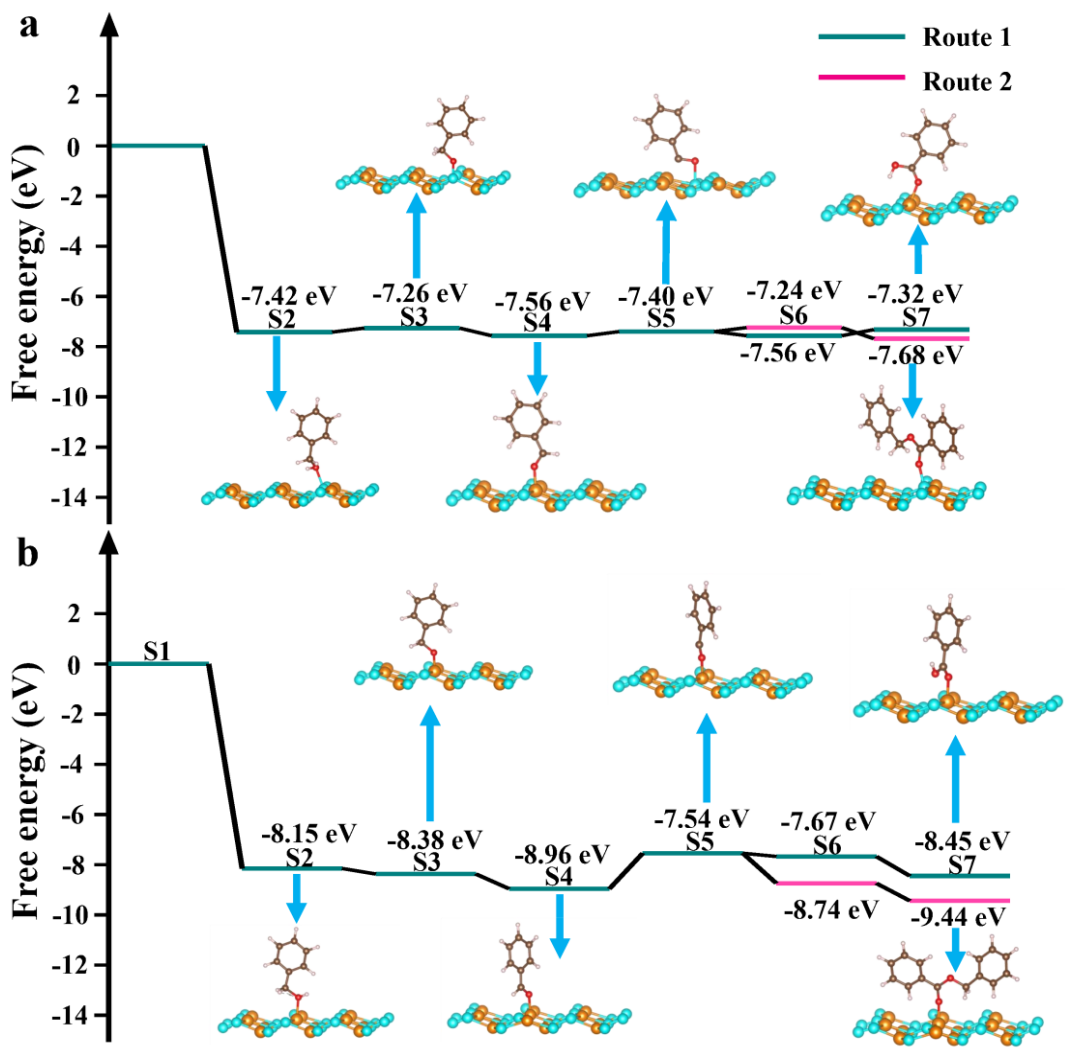

**Fig. 6.** Reaction energy profile for the formation of BAD and BBE on (200) crystal planes of S-CoSi (a) and CoSi$_v$ (b). Route 1: Conversion of BAL to BAD; Route 2: Conversion of BAL to BBE.

DFT calculations are also performed to explore the possible reaction pathway for the formation BAD and BBE on the S-CoSi, CoSi$_v$ (**Fig. 6**), and Co$_v$Si surfaces (Fig. S14), and the reaction energies of all intermediates are listed in Table S8 and Table S9. Firstly, BAL and oxygen are adsorbed on the surface, followed by cleavage of the O-H bond to form PhCH$_2$O* (* denotes the adsorbed species). Subsequently, the protons in PhCH$_2$O* are removed consecutively to form PhCHO* and then PhCO*. There are two reaction routes for PhCO*: PhCO* reacts with (1) PhCH$_2$O* to form BBE, or (2) OH to form BAD. On the CoSi$_v$ surface, the two steps of the BBE (BAD) formation reaction route are both exothermic with reaction energies: –1.21eV and –0.70eV (–0.13eV and –0.78eV). However, on the other two surfaces, the first step is exothermic, followed by the endothermic second step (with much larger reaction energies than those on the CoSi$_v$ surface). It thus indicates that BBE and BAD are more readily produced on the CoSi$_v$ surface. In the meantime, the reaction energies of the first step for forming BBE (–1.21eV) are significantly lower than BAD (–0.13eV), suggesting that BBE formation is the favored reaction route over BAD formation (the reaction energies –0.70eV and –0.78eV of the second steps are similar). Overall, our computational results are well matched with the catalytic results and demonstrate that Si vacancies in CoSi$_v$ with more Co sites on the surface are more catalytically reactive than S-CoSi, and Co$_v$Si surfaces (stronger reactants adsorption and exothermic reaction routes) for BBE formation.

## 4. Conclusions

In summary, semimetal CoSi alloy has been synthesized by the fast and solvent-free Arc-melting method in 5 mins. The developed CoSi alloy with divacancy exhibited a remarkable catalytic efficiency in the base- and solvent-free oxidation of six aromatic alcohols in comparison with its individual counterparts and the standard CoSi alloy, whereas 70% selectivity of BBE with up to 99.9% conversion of BAL was achieved. To the best of our knowledge, this is the highest BBE yield reported to date. Besides, these CoSi alloys maintain high stability even over 5 runs. In addition, these fabricated CoSi alloys also worked universally and exhibited high performance with over 84% selectivity for several types of aromatic alcohols with electron-donating and withdrawing groups (such as 2-, 3-, 4-methylbenzyl alcohol 4-methoxybenzyl alcohol and 4-nitrobenzyl alcohol). The integration of theory and experimental works confirmed that the superior performance and high stability of CoSi alloy are mainly attributed to the alloyed CoSi structure and Si vacancies. Semimetal CoSi alloys fabricated in this work by the facile and solvent-free method are efficient for the base- and solvent-free oxidation of various aromatic alcohols, which offers the rational design of other metal silicide alloy catalysts.

**Acknowledgments**

We acknowledge the National Center for High Performance Computing, Hsinchu, Taiwan for providing computational facilities. The calculation work is done by Fengyen Shi and Prof. Dr. Shi-Hsin Lin.

**Electronic supporting information**

Supporting information is available in the online version of this article.

# 富空穴CoSi合金无溶剂无碱醇氧化

赵志月[a,†], 蒋志伟[a,†], 黄一哲[a], Mebrouka Boubeche[b], Valentina G. Matveeva[c], Hector F. Garces[d], 罗惠霞[b], 严凯[a,e,*]

[a] 中山大学环境科学与工程学院，广东广州 510006
[b] 中山大学材料与工程学院，光电材料与技术国家重点实验室，广东广州 510006
[c] 特维尔国立技术大学生物技术与化学系，特维尔州特维尔市 170026，俄罗斯
[d] 布朗大学工程学院，罗德岛州普罗维登斯市 02912，美国
[e] 华南农业大学岭南现代农业科学与技术广东省实验室，广东广州 510642

**摘要：** 醇类化合物选择性氧化为醛、酸和酯在工业应用中是一类重要的化学反应。按照绿色化学的要求，无溶剂、无外加碱源，以分子氧为氧化剂，对醇类氧化反应提出了重要挑战，特别是无溶剂氧化醇制备酯类化学品。因此开发高效催化剂应用于无溶剂醇氧化得到学界的广泛关注。目前Pd基贵金属催化剂应用于醇的无溶剂氧化已取得了一定的进展，但是贵金属的稀缺性限制了其工业应用。此外，Co基非贵金属催化剂可以有效无溶剂氧化醇，但是其催化剂稳定性一般。因此，高效合成高活性高稳定性催化剂用于醇类化合物的无溶剂选择性氧化是一个重要的科学问题。

本文采用电弧熔炼法（AM）可控制备半金属CoSi合金（AM-CoSi），并将其应用于六种芳香醇的无溶剂无外加碱分子氧的选择性氧化。我们使用X射线吸收精细结构谱（XAFS）、电子顺磁共振（EPR）、球差扫描透射电镜（AC STEM）对AM-CoSi催化剂进行结构表征，发现其具有Co空穴和Si空穴，并且以Si空穴为主。催化结果表明，使用AM-CoSi催化剂在220 °C可完全转化苯甲醇，并可以得到产率为70%的苯甲酸苄酯（BBE），然而使用无缺陷CoSi合金，只得到9.5%的BBE。说明Si空穴存在可以高效的转化苯甲醇并得到高产率的BBE。此外，通过调控反应温度，使用AM-CoSi无溶剂氧化芳香醇可高选择性得到对应的醛，选择性高达84%以上。实验和DFT计算结果表明CoSi合金结构促进醇的转化，Si空穴有利于BBE的产生。本工作将为其他半金属合金催化剂的合成和应用提供新思路。

**关键词：** 半金属合金; 醇氧化; 无溶剂; 空穴; DFT计算

**Graphical Abstract**

# Facile synthesis of CoSi alloy with rich-vacancy for base- and solvent-free aerobic oxidation of aromatic alcohols

Zhiyue Zhao, Zhiwei Jiang, Yizhe Huang, Mebrouka Boubeche, Valentina G. Matveeva, Hector F. Garces, Huixia Luo, Kai Yan*

Sun Yat-Sen University; Tver State Technical University; Brown University; South China Agricultural University

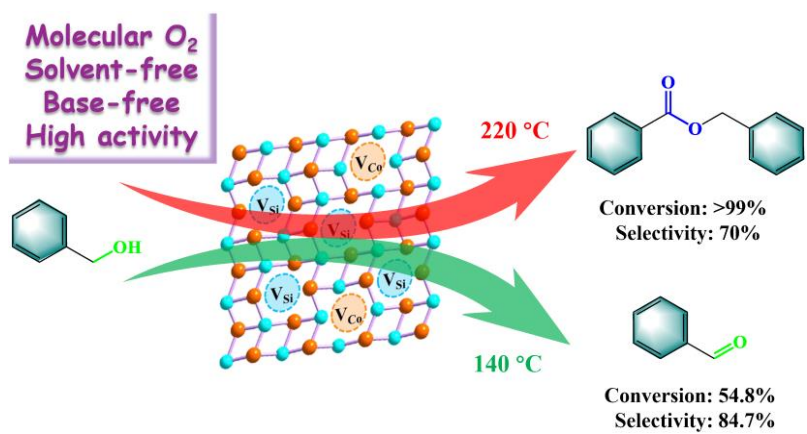

CoSi semimetal alloy with rich-vacancy is fast synthesized through arc-melting method, which can be bused as a high-efficiency catalyst for the the base- and solvent-free selective oxidation of various alcohols to aldehydes and esters.